\documentclass[seceq]{jpsj2} 
\usepackage{color}
\newcommand{\mvec}[1]{\textrm{\boldmath $#1$}}

\title{
On-line Learning of an Unlearnable True Teacher through Mobile Ensemble
Teachers
}

\author{Takeshi \textsc{Hirama} and Koji \textsc{Hukushima}}

\inst{Department of Basic Science, University of Tokyo, 3-8-1 Komaba, Meguro-ku, Tokyo 153-8902, Japan}
\recdate{\today}

\abst{
On-line learning of a hierarchical learning model is studied by a method
from statistical mechanics. In our model a student of a simple
perceptron learns from not a true teacher directly, but ensemble
teachers who learn from the true teacher with a perceptron learning
rule. 
Since the true teacher and the ensemble teachers are expressed as
non-monotonic perceptron and simple ones, respectively, the ensemble
teachers go around the unlearnable true teacher with the distance
between them fixed in an asymptotic steady state. 
The generalization performance of the student is shown to exceed that of
the ensemble teachers in a transient state, as was shown in similar
ensemble-teachers models. Further, it is found that moving the ensemble
teachers even in the steady state, in contrast to the fixed ensemble
teachers, is efficient for the performance of the student. 
}

\kword{online learning, ensemble teachers, generalization error,
statistical mechanics}

\begin{document}
\maketitle

\section{Introduction}
Learning is an inference problem of inhered rules from a given set of
examples which consist of input data and corresponding output data
generated by the rules. 
In practice, the examples are often supplied
inexhaustibly and then the learning must proceed by using each example
just once.   Such learning is called on-line learning\cite{2,3,1}. On
the contrary, the learning in which all the examples are presented
repeatedly at anytime is called off-line or batch learning. 

The on-line learning as well as the off-line one has
been extensively studied by using statistical-mechanical methods so
far and many extensions of the on-line
learning scheme have been made in order to improve a generalization
performance.\cite{1,2} Recently, 
Miyoshi and Okada \cite{6} and Urakami, Miyoshi and Okada \cite{4} analyzed 
the generalization performance of a student supervised 
by a moving teacher that goes around a fixed true teacher in a framework
of the on-line learning using the  statistical mechanical method. 
In their model, the student is not directly given the outputs by the
true teacher. The moving teacher learns from the true teacher and provides its
output to the student.  In this sense, the model is a kind of
hierarchical learning. In ref.~\citen{4}, 
the true teacher is a non-monotonic perceptron, while the moving teacher
and the student are simple perceptron using perceptron learning, 
which could not infer the true teacher completely in principle. 
The theoretical bound of the generalization error of a simple perceptron
learner has been obtained.\cite{InoueNishimori} 
In that case, the moving teacher goes around the true teacher with a
fixed distance between them. 
Interestingly, it turned out  that 
when the student's learning rate is relatively small, 
the student's generalization error can temporally become smaller than that of 
the moving teacher, even if the student only uses the examples from the
moving teacher.  

Subsequently, Miyoshi and Okada \cite{7} and Utsumi, Miyoshi and Okada
\cite{5} analyzed the generalization performance of 
an extended model of the on-line learning with 
multiple teachers, which would be called
ensemble-teachers learning model. This model is also regarded as an extension
of the ensemble learning\cite{Urbanczik,Miyoshi05} because the ensemble
teachers and the student in the ensemble-teachers model can be
interpreted as the ensemble students and their integrating mechanism,
respectively.  
In particular, ref.~\citen{5} discussed the model in which the true
teacher, the ensemble teachers and the student  are all simple perceptrons.
In this model the true teacher and the ensemble teachers are fixed. The
student adopts the Hebbian learning or the perceptron learning as a
learning rule and uses examples from the ensemble  
teachers in turn or randomly. As a result, it was clarified that the
Hebbian learning and the perceptron learning show qualitatively
different behavior from each other. In the Hebbian learning, the generalization
error monotonically decreases during the learning process and its
asymptotic value is independent of the learning rate. 
The asymptotic value is reduced as the number of the ensemble teachers
increases since the ensemble teachers have more variety in their
representations.  On the other hand, in the perceptron learning, the
generalization error shows non-monotonic behavior and exhibits a minimum
at a certain step in the learning. The minimum value of the
generalization error decreases as the learning rate decreases and the
total number of the teachers increases. 

In ref.~\citen{4} and ref.~\citen{5}, it was shown that the
generalization error of a student could be smaller than that of a moving
teacher or fixed ensemble teachers. 
A comparison between the generalization performance with a
fixed teacher and that with a mobile teacher, however, has not been made
directly. 
Furthermore, 
in the on-line learning with the ensemble teachers it is not
trivial that either the mobility or the multiplicity of the ensemble teachers is
effective for the learning performance of the student.   
In this paper, we study the on-line learning 
for the ensemble teachers which can move around a true teacher. 
We discuss a model in which the fixed true teacher is 
non-monotonic perceptron and the ensemble moving teachers and 
the student are a simple perceptron. 
This is a generalized version of the model studied in ref.~\citen{4}.
Adopting the   
perceptron learning as a learning rule for the ensemble teachers, they 
go around the true teacher with constant order parameters in the steady state. 
Then we analyze the generalization performance of the
student which learns from the mobile ensemble teachers using the Hebbian
and the perceptron rules. We also study the model with the ensemble
teachers fixed in their steady state.  It is thus clarified that the
movement of the ensemble teachers , in comparison with the fixed
ensemble case, significantly improves the generalization performance of
the student as a transient state in the learning process.  

The paper is organized as follows: 
In sec. $2$, we introduce the model with the ensemble moving teachers
going around the unlearnable true teacher.  
In sec. $3$, based on the statistical-mechanical idea, we theoretically 
derive the ordinal differential equations of order parameters and an explicit
formula of the generalization error of our model in terms of the order
parameters.  In sec. $4$, we show the theoretical and numerical results of 
the generalization performance of the student with the Hebbian and
perceptron rules. 
The last section is devoted to our conclusion. In the appendixes, 
the derivations of the differential equations discussed in
sec.~3 are presented in detail.

\section{Model}
In this paper, we consider a true teacher, $K$ ensemble moving teachers
and a student, whose connection weights are  expressed as $N$
dimensional vectors, \textrm{\boldmath $A$}, \textrm{\boldmath
$B$}$_{k}$ and \textrm{\boldmath $J$}, respectively, with
$k=1,2,\cdots,K$. 
For simplicity, each component $A_{i}$ of \textrm{\boldmath $A$} with
$i=1,\cdots, N$ is assumed to be drawn from $\mathcal{N}(0,1)$
independently and fixed, where $\mathcal{N}(m,\sigma^2)$ denotes the
Gaussian distribution with $m$ and $\sigma^2$ being a mean and variance,
respectively. 
As an initial condition of the learning process,  
each of the components $B_{ki}^{0}$ and $J_{i}^{0}$ of \textrm{\boldmath
$B$}$_{k}^0$, \textrm{\boldmath $J$}$^0$ are also assumed to be drawn from
$\mathcal{N}(0,1)$ independently. 
Input $\textrm{\boldmath $x$}$ is also the $N$-dimensional vector and the
component $x_{i}$  follows from $\mathcal{N}(0,1/N)$ independently. 
Thus, we have
\begin{equation}
 \left\langle A_i \right\rangle = \left\langle B_{ki}^0 \right\rangle = 
\left\langle J_{i}^{0} \right\rangle = \left\langle x_i \right\rangle = 0,
\end{equation}
\begin{equation}
\left\langle (A_i)^2 \right\rangle = 
\left\langle \left(B_{ki}^0\right)^2 \right\rangle = 
\left\langle \left(J_{i}^{0}\right)^2 \right\rangle =1, 
\end{equation}
and 
\begin{equation}
\left\langle (x_i)^2 \right\rangle =\frac{1}{N}, 
\end{equation}
where $\langle \cdots \rangle$ denotes an average over the Gaussian
distribution. 

In the statistical mechanics of the learning,\cite{2,3} we are interested in
asymptotic behavior of $\mvec{A}$, \mvec{B} and \mvec{J} in a
thermodynamics limit $N\to \infty$. 
Then, one finds that the norms of the vectors are 
\begin{equation}
\Vert \textrm{\boldmath $A$}\Vert=\sqrt{N}, \,\Vert \textrm{\boldmath $B$}_{k}^0 \Vert =\sqrt{N}, \, \Vert \textrm{\boldmath $J$}^0 \Vert =\sqrt{N}, \, \Vert \textrm{\boldmath $x$} \Vert=1.
\end{equation}
The norms, 
$\Vert \textrm{\boldmath $B$}_k \Vert$ and $\Vert \textrm{\boldmath $J$}
\Vert$, of the ensemble moving teachers and the student change during
the learning process from their initial values. 
The normalized length of these vectors is introduced as $l_{B_k}=\Vert
\textrm{\boldmath $B$ }_k\Vert/\Vert \textrm{\boldmath$B$}_k^{0}\Vert$ for the ensemble
teachers and $l_{J}=\Vert\textrm{\boldmath$J$}\Vert/\Vert\textrm{\boldmath$J$}^{0}\Vert$
for the student. 
In the thermodynamic limit, the direction cosines between these vectors
are a relevant extensive quantity, denoted for \textrm{\boldmath $A$} and
\textrm{\boldmath $B$}$_{k}$, 
\textrm{\boldmath $A$} and \textrm{\boldmath $J$}, 
 \textrm{\boldmath $B$}$_{k}$ and \textrm{\boldmath $B$}$_{k'}$, 
and  \textrm{\boldmath $B$}$_{k}$ and \textrm{\boldmath $J$} respectively as
\begin{align}
R_{B_k}&=\frac{\textrm{\boldmath $A$}\cdot\textrm{\boldmath $B$}_k}{\Vert \textrm{\boldmath $A$}\Vert \Vert\textrm{\boldmath $B$}_k \Vert}, \, R_{J}=\frac{\textrm{\boldmath $A$}\cdot\textrm{\boldmath $J$}}{\Vert \textrm{\boldmath $A$}\Vert \Vert\textrm{\boldmath $J$} \Vert},   \\
q_{kk'}&=\frac{\textrm{\boldmath $B$}_k\cdot\textrm{\boldmath
 $B$}_k'}{\Vert \textrm{\boldmath $B$}_k\Vert \Vert\textrm{\boldmath
 $B$}_k' \Vert}, \, R_{B_{k}J}=\frac{\textrm{\boldmath
 $B$}_{k}\cdot\textrm{\boldmath $J$}}{\Vert \textrm{\boldmath
 $B$}_{k}\Vert \Vert\textrm{\boldmath $J$} \Vert}.    
\end{align}

In the present study, we assume that the true teacher is a non-monotonic
perceptron and the ensemble moving teachers and the student are a
simple perceptron. 
The output for a given input \mvec{x} of the true teacher is defined by a non-monotonic function 
\begin{equation}
o=\text{sgn}\left(\left(\textrm{\boldmath $A$}\cdot\textrm{\boldmath $x$}-a\right)\textrm{\boldmath $A$}\cdot\textrm{\boldmath $x$}\left(\textrm{\boldmath $A$}\cdot\textrm{\boldmath $x$}+a\right)\right)
\end{equation}
with a fixed threshold $a$,   
while those of the ensemble moving teachers and the student
are simply given by sgn$\left(\textrm{\boldmath $B$}_k\cdot\textrm{\boldmath $x$}\right)$ and
sgn$\left(\textrm{\boldmath $J$}\cdot\textrm{\boldmath $x$}\right)$, respectively. Here, sgn$(\cdot)$ is
the sign function defined as 
\begin{equation}
\text{sgn}(s)=\left\{
\begin{array}{ll}
 +1, &\quad s\geq 0,  \\
 -1, &\quad s < 0.
\end{array}
\right.
\end{equation}
A measure of dissimilarity between the true teacher and the ensemble
teachers or the student is defined by using their outputs  as 
\begin{equation}
\epsilon_{B_{k}}\equiv\Theta\left(-o\cdot\text{sgn}\left(\textrm{\boldmath $B$}_k\cdot\textrm{\boldmath $x$}\right)\right) 
\label{eqn:errorB}
\end{equation}
for $k$th ensemble teacher and 
\begin{equation}
\epsilon_{J}\equiv\Theta\left(-o\cdot\text{sgn}\left(\textrm{\boldmath $J$}\cdot\textrm{\boldmath $x$}\right)\right)
\label{eqn:errorJ}
\end{equation}
for the student, where $\Theta(\cdot)$ is the step function defined as 
\begin{equation}
\Theta(s)=\left\{
\begin{array}{ll}
 +1, &\quad s\geq 0,  \\
 \,\,\,\,0, &\quad s < 0.
\end{array}
\right.
\end{equation}
One of the main purposes of the statistical learning theory is to obtain
theoretically the generalization errors $\epsilon_{B_{k}}^{g}$ and
$\epsilon_{J}^{g}$, which are defined as the average of the errors, 
$\epsilon_{B_{k}}$ and $\epsilon_{J}$ over the whole set of possible
inputs $\textrm{\boldmath $x$}$. 
Since the input \mvec{x} appears in
Eq.~(\ref{eqn:errorB}) and Eq.~(\ref{eqn:errorJ}) 
as inner products $\mvec{A}\cdots\mvec{x}$, $\mvec{B}_k\cdot\mvec{x}$
and $\mvec{J}\cdot\mvec{x}$, the average over Gaussian vector $\mvec{x}$
could be reduced to an average over correlated Gaussian variables. When
one defines  a set of variables, $v$, $v_{B_{k}}$ and $u$ as 
\begin{align}
v&= \textrm{\boldmath $A$}\cdot \textrm{\boldmath $x$},\\
v_{B_{k}}l_{B_{k}}&= \textrm{\boldmath $B$}_{k}\cdot \textrm{\boldmath $x$},\\
ul_{J}&= \textrm{\boldmath $J$}\cdot \textrm{\boldmath $x$},
\end{align}
they obey the multiple Gaussian distribution 
\begin{equation}
P(v, \{v_{B_{k}}\}, u) =\frac{1}{(2\pi)^{(K+2)/2}\vert \Sigma \vert
 ^{1/2}}\exp{\left(-\frac{(v, \{v_{B_{k}}\}, u)\Sigma^{-1}(v,
	      \{v_{B_{k}}\}, u)^{T}}{2}\right)},
\label{eqn:multiG}
\end{equation}
with zero means and the covariance matrix $\Sigma$
\begin{align}
\Sigma &=\left(
\begin{array}{cccccc}
1 & R_{B_{1}}& R_{B_{2}}& \cdots & R_{B_{K}}& R_{J} \\
 R_{B_{1}}& 1& q_{1,2}&\cdots & q_{1,K}& R_{B_{1}J} \\
 R_{B_{2}}& q_{2,1}& 1& \ddots & \vdots& \vdots \\
 \vdots& \vdots& \ddots & \ddots& q_{K-1,K}& R_{B_{K-1}J} \\
 R_{B_{K}}& q_{K,1}& \cdots& q_{K,K-1}& 1& R_{B_{K}J}\\
 R_{J}&R_{B_{1}J} &\cdots &R_{B_{K-1}J} & R_{B_{K}J}& 1
\end{array}
\right).\label{covariance}
\end{align}
Evaluating the correlated Gaussian integrations, 
the generalization errors $\epsilon_{B_{k}}^{g}$ and
$\epsilon_{J}^{g}$ are obtained as  
\begin{align}
\epsilon_{B_{k}}^{g}&=2\left(\int_{-\infty}^{-a}+\int_{0}^{a}\right)Dv H\left(-\frac{R_{B_{k}}v}{\sqrt{1-R_{B_{k}}^2}}\right), \label{gerrorAB}
\end{align}
and 
\begin{align}
\epsilon_{J}^{g}&=2\left(\int_{-\infty}^{-a}+\int_{0}^{a}\right)Dv
 H\left(-\frac{R_{J}v}{\sqrt{1-R_{J}^2}}\right), \label{gerrorAJ}
\end{align}
where $Ds$ is the Gaussian measure defined as
\begin{equation}
Ds\equiv \frac{ds}{\sqrt{2\pi}}\exp{\left(-\frac{s^2}{2}\right)}, 
\end{equation}
and $H(\cdot)$ is the error function defined as
\begin{equation}
H(s)\equiv \int_{s}^{\infty} Dx. 
\end{equation}
It should be noted that the dynamical effect of the generalization
errors appears only through $R_{B_{k}}$ and $R_{J}$. 
This implies that the generalization errors have a
fundamental minimum as a function of $R_{B_{k}}$ and $R_{J}$,
irrespective of the matter if the values of $R_{B_{k}}$ and $R_{J}$
which give the minimum value of the generalization error appear in a
particular chosen learning rule of the student and the ensemble
teachers. An efficient learning rule might realize the fundamental
minimum for  a given learning model.

Let us defined the update rule in the on-line learning.  
The ensemble moving teachers \textrm{\boldmath $B$}$_{k}$ are updated
from the current state \textrm{\boldmath $B$}$_k^{m'}$
using an input $\textrm{\boldmath $x$}$ and output of the true teacher
\textrm{\boldmath $A$} for the input  \textrm{\boldmath $x$}$^{m'}$,
independently as 
\begin{align}
\textrm{\boldmath $B$}_{k}^{m'+1}= \textrm{\boldmath $B$}_{k}^{m'}+f_{k}^{m'}
(\textrm{\boldmath $x$}^{m'}, \textrm{\boldmath $B$}_{k}^{m'},
 o^{m'})\textrm{\boldmath $x$}^{m'},  
\label{eqn:updateB}
\end{align}
where $f^{m'}$ is  an update function of the ensemble moving teachers and
$m'$ denotes the time step of the ensemble moving teachers. 
In particular, we choose the perceptron learning for the update function
$f_k$, which is given by 
\begin{align}
f_{k}^{m'}=\eta_{B}\Theta\left(-v_{B_{k}}^{m'}o^{m'}\right)o^{m'}. 
\end{align}
Here, $\eta_{B}$ is the learning rate of the ensemble moving teachers. 
In our analysis, the learning rate $\eta_{B}$ is independent of the
teachers and is fixed during the learning process. 
After a sufficient long learning process using the perceptron rule, 
the ensemble moving teachers reach steady state with 
$R_{B_{k}}$, $l_{B_{k}}$ and $q_{kk'}$ fixed. 
In the present study, we focus our attention to dynamical effect of the
ensemble teachers for the learning performance of the student. In order
to separate off a transient effect of the ensemble teachers, the student learns from the ensemble teachers in the steady state. 
The student \textrm{\boldmath $J$} is updated using 
an input  \textrm{\boldmath $x$} and an output of one of the $K$
ensemble moving teachers \textrm{\boldmath $B$}$_{k}$ chosen randomly. 
The explicit recursion formula for $\textrm{\boldmath $J$}^{m}$ with $m$
being the time step of the student is given by 
\begin{align}
\textrm{\boldmath $J$}^{m+1}= \textrm{\boldmath $J$}^{m}+g_{k}^{m}
(\textrm{\boldmath $x$}^{m}, \textrm{\boldmath $J$}^{m},
 \text{sgn}(v_{B_{k}}l_{B_{k}}))\textrm{\boldmath $x$}^{m},  
\label{eqn:updateJ}
\end{align}
where $g_{k}^{m}$ is an update function of the student and $k$ is a
uniform random integer chosen from $1$ to $K$.  Note that the
ensemble moving teachers are also updated using the same input. 
We particularly discuss two different learning rules for the student,
which are the Hebbian learning
\begin{equation}
g_{k}^{m}=\eta
 \text{sgn}\left(v_{B_{k}}^{m}l_{B_{k}}^{m}\right), 
\label{eqn:Hrule}
\end{equation}
and the perceptron learning 
\begin{equation}
g_{k}^{m}=\eta
 \Theta\left(-v_{B_{k}}^{m}u^{m}\right)\text{sgn}\left(v_{B_{k}}^{m}l_{B_{k}}^{m}\right). 
\label{eqn:Prule}
\end{equation} 
The learning rate of the student $\eta$ is also  constant 
during the learning process.

\section{Order-parameter theory}
As shown in the previous section, the generalization errors of the
ensemble teachers and the student are expressed in terms of the
parameter $R_{B_k}$ and $R_{J}$ and evolve only trough a few parameters
associated with the learning of $\textrm{\boldmath $B$}_{k}$ and
$\textrm{\boldmath $J$}$ in the thermodynamic limit. It has been shown
that a class of the on-line learning can be characterized by a few
extensive parameters, called order parameter. In this section, following ref.~\citen{1}, a set of ordinal differential equations of the order
parameters are obtained in our model by taking the thermodynamic limit. 

The learning process of the ensemble moving teachers are described by
the three order parameter $R_{B_k}$, $l_k$ and $q_{kk'}$, which are
assumed to be self-averaging. 
It is sufficient to consider the evolution of $R_{B_k}$ and $l_k$ in
order to describe the dynamics of the ensemble teachers, but that of the
overlap $q_{kk'}$ between two different teachers is necessary for the
student dynamics as seen later. 
From the update rules of the ensemble
teachers in eq.~(\ref{eqn:updateB}), one finds a closed formula of the
ordinal differential equations of the order parameters as, 

\begin{align}
\frac{dl_{B}}{dt'}&=
\frac{\eta_B}{\sqrt{2\pi}}
 \left[R_{B}\left\{2\exp{\left(-\frac{a^2}{2}\right)}-1\right\}-1\right]+
 \frac{1}{2l_B}\left(
2\eta_B^2\left(\int_{-\infty}^{-a}+\int_{0}^{a}\right)Dv H\left(-\frac{R_{B}v}{\sqrt{1-R_{B}^2}}\right) 
\right)
, \label{eqn:opeL} \\
\frac{dR_{B}}{dt'}&=
-\frac{R_{B}}{l_{B}}\frac{dl_{B}}{dt'}+\frac{1}{l_{B}}\left(
\frac{\eta_B}{\sqrt{2\pi}}\left\{2\exp{\left(-\frac{a^2}{2}\right)}-R_{B}-1\right\}
\right)
, \label{eqn:opeR} \\
\frac{dq}{dt'}&=
-\frac{q}{l_{B}}\frac{dl_{B}}{dt'}-\frac{q}{l_{B}}\frac{dl_{B}}{dt'} 
+\frac{2}{ l_{B}}\left(
\frac{\eta_B}{\sqrt{2\pi}}\left[R_{B}\left\{2\exp{\left(-\frac{a^2}{2}\right)}-1\right\}-q\right]
\right)\nonumber \\
&
+\frac{1}{l_{B}^2}\left(
2\eta_{B}^2\left(\int_{-\infty}^{-a}+\int_{0}^{a}\right)Dv\int_{-\frac{R_{B}v}{\sqrt{1-R_{B}^2}}}^{\infty}DxH(z)
\right), \label{eqn:opeQ} 
\end{align}
where 
\begin{equation}
z\equiv-\frac{(q-R_{B}^2)x+R_{B}\sqrt{1-R_{B}^2}v}{\sqrt{(1-q)(1+q-2R_{B}^2)}}, 
\end{equation}
and $t'$ denotes continuous time. We omit the subscript $k$ from the
order parameters, because the differential equations including their
initial conditions have a permutation symmetry for the subscript
$k$. Derivation of the differential equation is given in the appendix\ref{sec:A}.

From these equation one easily obtain the steady solutions of $R_{B}$,
$l_{B}$ and of $q$ as follows: 
\begin{align}
R_{B}&=2\exp{\left(-\frac{a^2}{2}\right)}-1,\\
l_{B}&=\frac{\sqrt{2\pi}\eta_{B}}{1-R_{B}^2}\left(\int_{-\infty}^{-a}+\int_{0}^{a}\right)Dv H\left(-\frac{R_{B}v}{\sqrt{1-R_{B}^2}}\right), \\
q&=R_{B}^2+\frac{\displaystyle (1-R_{B}^2)\left(\int_{-\infty}^{-a}+\int_{0}^{a}\right)Dv\int_{-\frac{R_{B}v}{\sqrt{1-R_{B}^2}}}^{\infty}DxH(z)}{\displaystyle\left(\int_{-\infty}^{-a}+\int_{0}^{a}\right)Dv H\left(-\frac{R_{B}v}{\sqrt{1-R_{B}^2}}\right)}.
\end{align}
Note that $R_{B}$, $q$ and  $l_{B}/\eta_{B}$ depend only on the
threshold $a$ of the true teacher.
In our study, the ensemble teachers are assumed to take the steady state
before the student begins to learn in order to make the
dynamical effect of the ensemble teachers clear. Therefore these
solutions of $R_B$, $l_B$ and $q$ are used as an initial condition of
the learning dynamics of the student discussed below.

The learning dynamics of the student is also described by a set of 
ordinal differential equations of a few order parameters, which is
derived from the update functions for the Hebbian rule
(\ref{eqn:Hrule}) and the perceptron one (\ref{eqn:Prule}).  We refer to
the appendix\ref{sec:B} for the derivation of the 
dynamical equations. A straightforward calculation for the Hebbian rule
leads to 
\begin{eqnarray}
 \frac{dl}{dt} & = & \eta\left(\sqrt{\frac{2}{\pi}}R_B+\frac{\eta}{2l}\right), \label{eqn:opeSH1}\\
 \frac{dR_J}{dt} & = & -\frac{R_J}{l}\frac{dl}{dt}+\frac{\eta}{l}\sqrt{\frac{2}{\pi}}R_B, \label{eqn:opeSH2}\\
 \frac{dR_{BJ}}{dt} & = &-R_{BJ}\left(\frac{1}{l}\frac{dl}{dt}+\frac{1}{l_B}\frac{dl_B}{dt}\right)
+
\frac{\eta_B}{l_B\sqrt{2\pi}}\left\{R_J\left(2e^{-\frac{a^2}{2}}-1\right)-R_{BJ}\right\}\nonumber \\
&+&\frac{\eta}{lK}\sqrt{\frac{2}{\pi}}q -\frac{2\eta\eta_B}{Kl_Bl}\left(
(K-1)\left(\int_{-\infty}^{-a}+\int_{0}^{a}\right)Dv
      \int_{-\frac{R_{B}v}{\sqrt{1-R_{B}^2}}}^{\infty} Dx
      \left\{2H(z)-1\right\} \right. \nonumber \\
 &+& \left.\left(\int_{-\infty}^{-a}+\int_{0}^{a}\right)Dv
 H\left(-\frac{R_{B}v}{\sqrt{1-R_{B}^2}}\right)
\right)
\label{eqn:opeSH3}. 
\end{eqnarray}
Corresponding differential equations for the perceptron rule are given
as 
\begin{eqnarray}
 \frac{dl}{dt} & = & \eta\left(\frac{R_{BJ}-1}{\sqrt{2\pi}}+\frac{\eta}{\pi}\tan^{-1}\left(\frac{\sqrt{1-R_{BJ}^2}}{R_{BJ}}\right)\right),\label{eqn:opeSP1}\\
 \frac{dR_J}{dt} & = & -\frac{R_J}{l}\frac{dl}{dt}+\frac{\eta}{l\sqrt{2\pi}}(R_B-R_J), \label{eqn:opeSP2}\\
 \frac{dR_{BJ}}{dt} & = & -R_{BJ}\left(\frac{1}{l}\frac{dl}{dt}+\frac{1}{l_B}\frac{dl_B}{dt}\right)+
\frac{\eta_B}{l_B\sqrt{2\pi}}\left\{R_J\left(2e^{-\frac{a^2}{2}}-1\right)-R_{BJ}\right\}\nonumber \\
&+&\frac{\eta  q}{lK\sqrt{2\pi}}\left(\frac{q}{K}-R_{BJ}\right)\nonumber \\
&+&\frac{2\eta\eta_B}{Kl_Bl}\left((K-1)
\left(\int_{-\infty}^{-a}+\int_{0}^{a}\right)Dv \int_{-\frac{R_{B}v}{\sqrt{1-R_{B}^2}}}^{\infty} Dx \left\{-\int_{z}^{\infty}DyH\left(-z_1\right)+\int_{-\infty}^{z}DyH\left(z_1\right)\right\}
\right. \nonumber \\
&+& \left.\left(\int_{-\infty}^{-a}+\int_{0}^{a}\right)Dv
 \int_{-\frac{R_{B}v}{\sqrt{1-R_{B}^2}}}^{\infty} Dx
 \left\{2H(z_2)-1\right\}\right)
 \label{eqn:opeSP3}
\end{eqnarray}
Solving these differential equations for the student and the ensemble
teachers, we can obtain the generalization errors $\epsilon_J^g$ and
$R_J$ as a function of time step. 


\section{Results and Discussion}

In this section we present dynamical behavior of the order parameter
$R_J$ and the generalization error $\epsilon_J^g$ obtained by solving
numerically the set of the differential equations obtained in the
previous section. In order to study
``dynamical'' effect of the ensemble teachers, we compare results of two
different cases; one with the teachers fixed to a steady state and the
other with the teachers kept to learn in the steady state sharing the
same inputs with the student. In this study, we choose the threshold value
$a=0.5$ of the non-monotonic perceptron for the true teacher, yielding 
$l_{B}/\eta_{B}\simeq 0.93$, $R_{B}\simeq 0.76$ and $q\simeq 0.91$ in
the steady state for the ensemble teachers. We also perform direct
simulations of the given update rules for the finite-size
perceptrons. In the simulations we use the dimension of vectors $N=10^4$
and perform $10^5$ trajectories of the learning process for taking the
average over the random inputs. As shown in figures below, 
although a limited case with $\eta=0.1$ is only shown for avoiding
crowded plots, the results of $R_J$ and $\epsilon_J^g$ obtained by the
simulations for all the parameter studied  agree with the theoretical
ones by the order-parameter differential equations, 
This confirms that the assumption of the self-averaging is appropriate in our
model. 

\begin{figure}[h]
 \begin{center}
    \resizebox{0.5\textwidth}{!}{\includegraphics{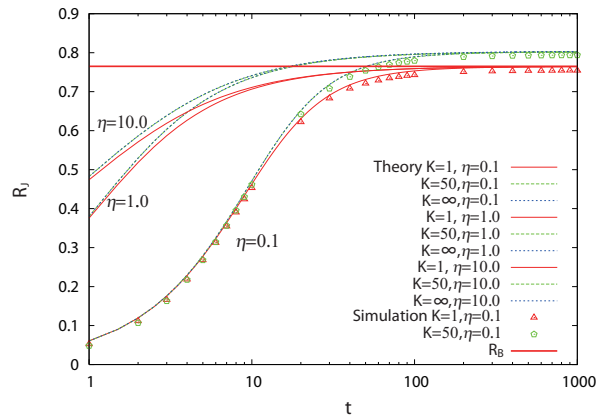}}
     \caption{
Time dependence of the direction cosine $R_{J}$ between the student \mvec{J}
 with the Hebbian learning and the true teacher \mvec{A}
    with $a=0.5$ 
 in the case that the $K$ ensemble teachers are fixed to be a steady state
  vector. 
Curves represent numerical solution of the order-parameter differential
 equations with $K=1, 50$ and $\infty$ and $\eta=0.1, 1$ and
 $10$. 
The straight line is the direction cosine $R_B$ between the fixed
  ensemble teachers and the true teacher. 
Symbols represent corresponding results obtained by 
    the direct simulation with system size $N=10^4$ and $\eta=0.1$. 
}
    \label{f1}
\end{center}
\end{figure}

\begin{figure}
\begin{center}
    \resizebox{0.5\textwidth}{!}{\includegraphics{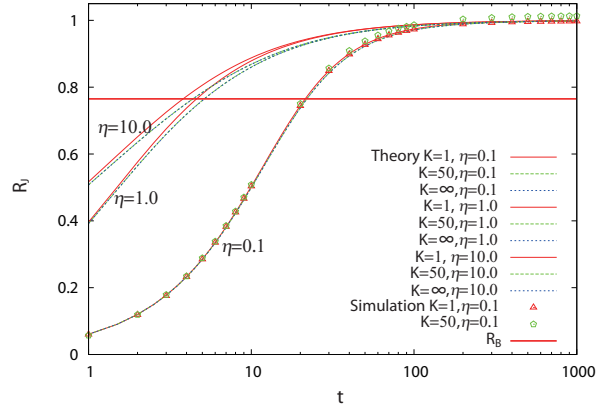}}
     \caption{
Time dependence of the direction cosine $R_{J}$ between the student \mvec{J}
 with the Hebbian learning and the true teacher \mvec{A}
    with $a=0.5$
 in the case that the ensemble teachers continue to learn in their steady state. 
The symbols and the lines are the same as those in Fig.~\ref{f1}. 
}

     \label{f2}
 \end{center}
\end{figure}

Figure~\ref{f1} shows time dependence of $R_J$ for the Hebbian learning
when the ensemble teachers stop to learn and take a steady-state
vector. 
The transient process of $R_J$ depends on the learning
rate $\eta$ of the student and the number $K$ of the ensemble teachers.
The value of $R_J$ gets larger with increasing the number $K$ and the
learning rate $\eta$, meaning that the student comes close to the true
teacher. As the time $t$ goes on, it approaches monotonically a steady
value, which increases as $K$  increases. Interestingly,  the steady
value of $R_J$ exceeds the value of 
$R_B$ when the number $K$ of  the ensemble teachers is greater than 1. 
This is similar to that shown in ref.~\citen{5}. 
Figure \ref{f2}, on the other hand,  shows the corresponding time
dependence of $R_J$ when the ensemble teachers continue to learn in
their steady state. While at the very beginning of the learning process
the value of $R_J$  shows monotonic time development similar to the case
that the ensemble teachers are fixed, it is larger than that with the
fixed teachers after a certain time and eventually approaches unity,
which is independent of the learning rate, even if the number $K$  is
one. It should be noted that the value of $R_B$ is common in two cases of
Figs. \ref{f1} and \ref{f2}. 
This implies that the number $K$ of the ensemble teachers is not
efficient for the learning of the student, but their continuous learning
even with a fixed similarity to the true teacher is significantly
important. 

\begin{figure}[h]
 \begin{center}
 \resizebox{0.5\textwidth}{!}{\includegraphics{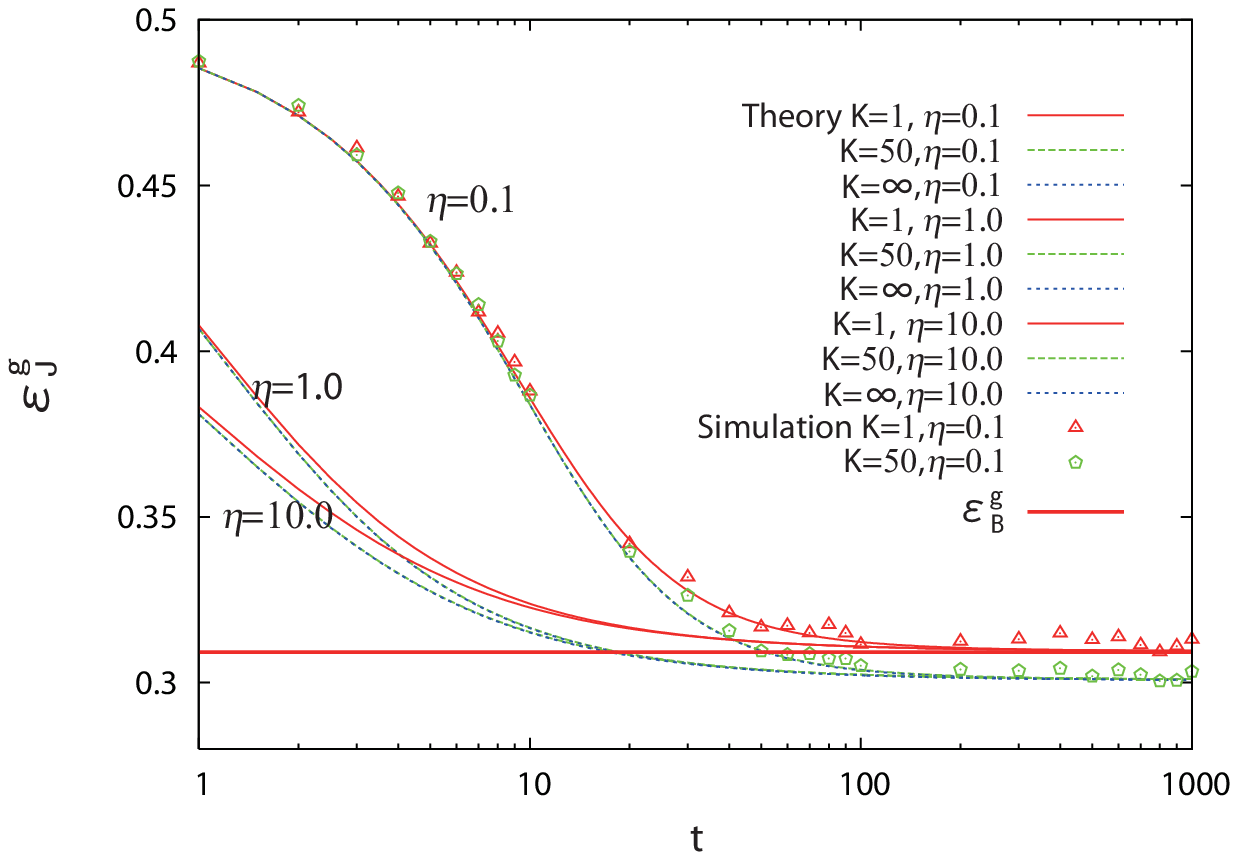}}
     \caption{Time dependence of the generalization error of the student $\epsilon_{J}$ between the student \mvec{J}
 with the Hebbian learning and the true teacher \mvec{A}
    with $a=0.5$  in the case of  the fixed ensemble teachers. 
The symbols and lines are the same as those in Fig.~\ref{f1}. 
}
 \label{f3}
 \end{center}
\end{figure}
\begin{figure}
 \begin{center}
 \resizebox{0.5\textwidth}{!}{\includegraphics{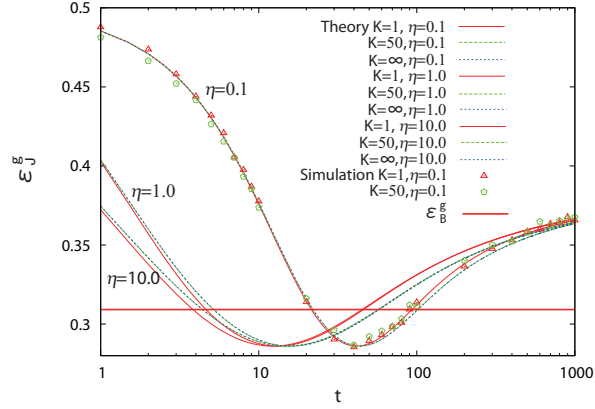}}
     \caption{Time dependence of the generalization error of the student
 $\epsilon_{J}$ between the student \mvec{J} with the Hebbian learning
 and the true teacher \mvec{A} with $a=0.5$ in the case of the mobile
 ensemble teachers. The symbols of the lines and plots are the same as
 in Fig.~\ref{f1}. 
}
     \label{f4}
\end{center}
\end{figure}

Figure~\ref{f3} shows dynamical behavior of the generalization error of
the student for the Hebbian learning, which 
monotonically decreases and eventually converges to the steady value
when the ensemble teachers are fixed. 
The steady value of $\epsilon_J^g$ only depends on the number $K$ and not
the learning rate $\eta$. As $K$ increases, the value decreases and
furthermore it can be smaller than that of the generalization error
$\epsilon_B^g$ of the ensemble teachers when $K$  is larger than one,
reflecting the behavior of $R_J$. 
This means that the performance of the student becomes better than the
ensemble teachers when $K\geq 2$. 
The obtained value of $\epsilon_J^g$, however, does not reach the
fundamental minimum value of the generalization error in this case even when $K$
increases to infinity. In Fig.~\ref{f4} the dynamical behavior of
$\epsilon_J^g$ is shown in the case where the ensemble teachers are
moving. In contrast to the case of the fixed ensemble teachers,
$\epsilon_J^g$ shows non-monotonic behavior in the learning process  and
the steady value of independent of both $K$ and $\eta$ while it is quite
larger than $\epsilon_B^g$. 
The minimum value of $\epsilon_J^g$ reaches the fundamental minimum
value at a certain time step, depending on the learning rate $\eta$. 
In a sense, the mobile ensemble teachers is a better on-line learning
model, while the best performance occurs only at a transient state
unfortunately. 

\begin{figure}[h]
 \begin{center}
  \resizebox{0.5\textwidth}{!}{\includegraphics{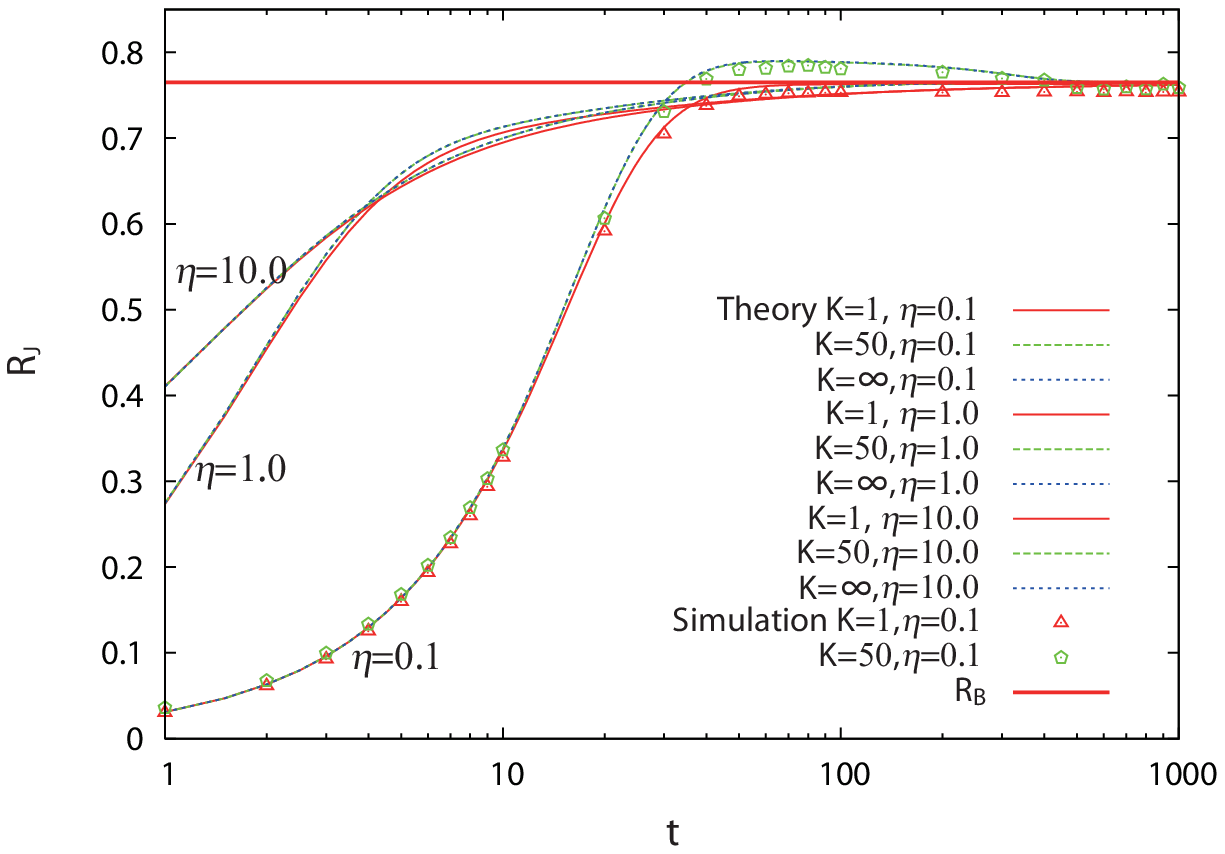}}
     \caption{Time dependence of the direction cosine $R_{J}$ between
  the student \mvec{J} with the perceptron learning and the true teacher
  \mvec{A} with $a=0.5$ in the case of the fixed ensemble teachers. 
The symbols of the lines and plots are the same as
 in Fig.~\ref{f1}. 
}
  \label{f5}
\end{center}
\end{figure}

\begin{figure}
 \begin{center}
  \resizebox{0.5\textwidth}{!}{\includegraphics{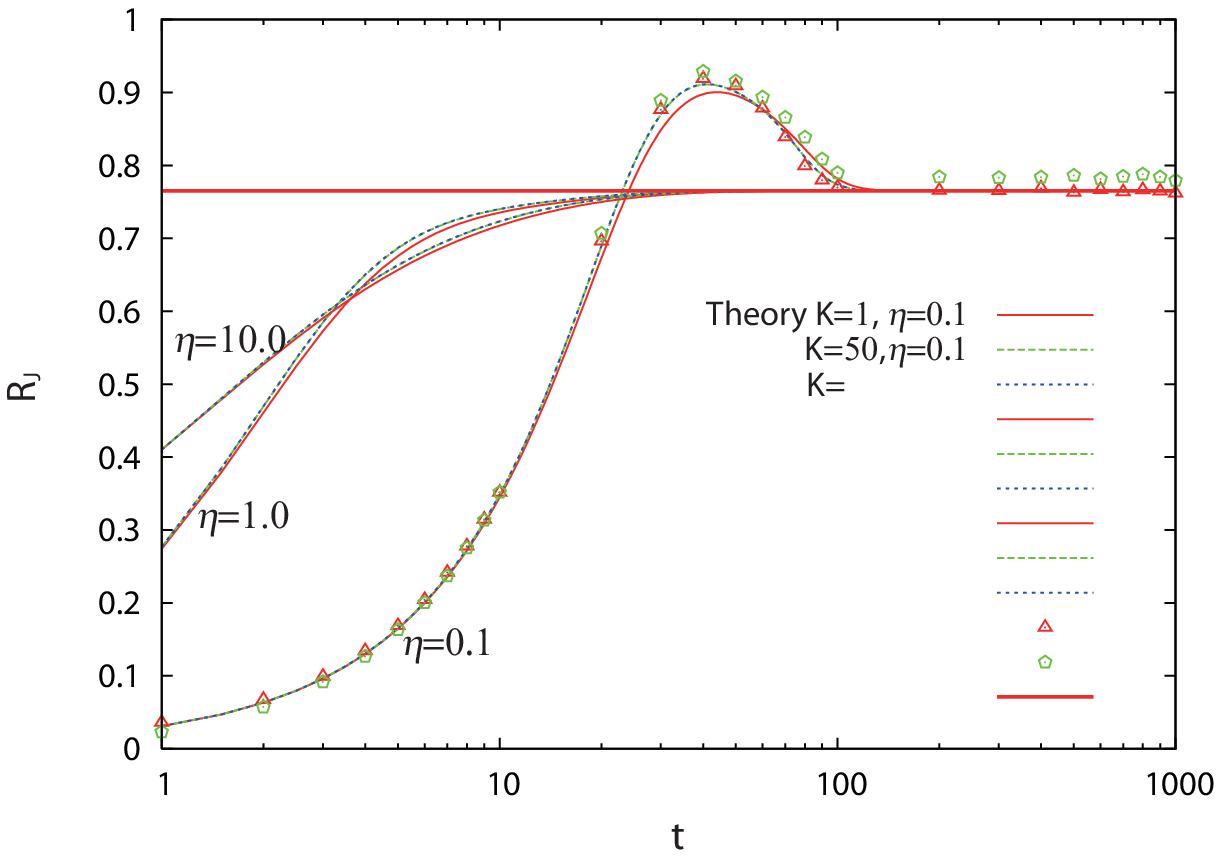}}
     \caption{Time dependence of the direction cosine $R_{J}$ between
  the student \mvec{J} with the perceptron learning and the true teacher
  \mvec{A} with $a=0.5$ in the case of the mobile ensemble teachers. 
The symbols of the lines and plots are the same as
 in Fig.~\ref{f1}. 
}
     \label{f6}
 \end{center}
\end{figure}

Let us turn to the perceptron learning of the student. 
We show the time development of $R_J$ for the fixed and moving ensemble
teachers in Figs.~\ref{f5} and \ref{f6}, respectively. The steady values 
of $R_J$ coincide with $R_B$ both for the two cases and it is
independent of $K$ and $\eta$. Further non-monotonic behavior is found
for small $\eta$ and large $K$ and then the value of $R_J$ takes a
maximum value at a certain time step, which exceeds $R_B$ certainly as a
transient state.  
Moving the ensemble teachers enhances significantly the maximum value,
meaning that the student is closer to the true teacher. In particular,
for small value of $\eta$ the maximum value of $R_J$ for the unique moving 
teacher is larger than that for the $K=\infty$ fixed ensemble teachers. 

\begin{figure}
 \begin{center}
    \resizebox{0.5\textwidth}{!}{\includegraphics{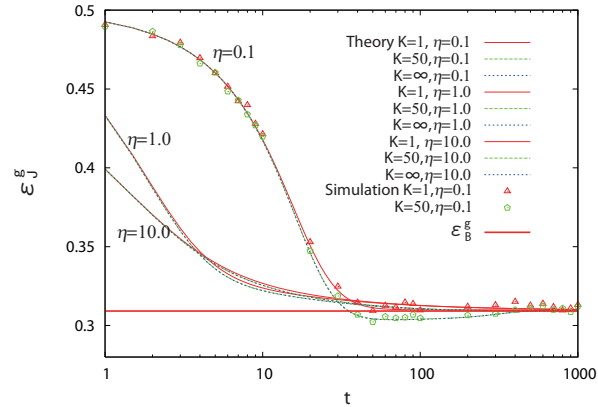}}
     \caption{Time dependence of the generalization error of the student
  $\epsilon_{J}$ between the student \mvec{J} with the perceptron
  learning and the true teacher \mvec{A} with $a=0.5$ in the case that
  the fixed ensemble teachers. 
The symbols of the lines and plots are the same as
 in Fig.~\ref{f1}. 
}
    \label{f7}
\end{center}
\end{figure}

\begin{figure}
 \begin{center}
  \resizebox{0.5\textwidth}{!}{\includegraphics{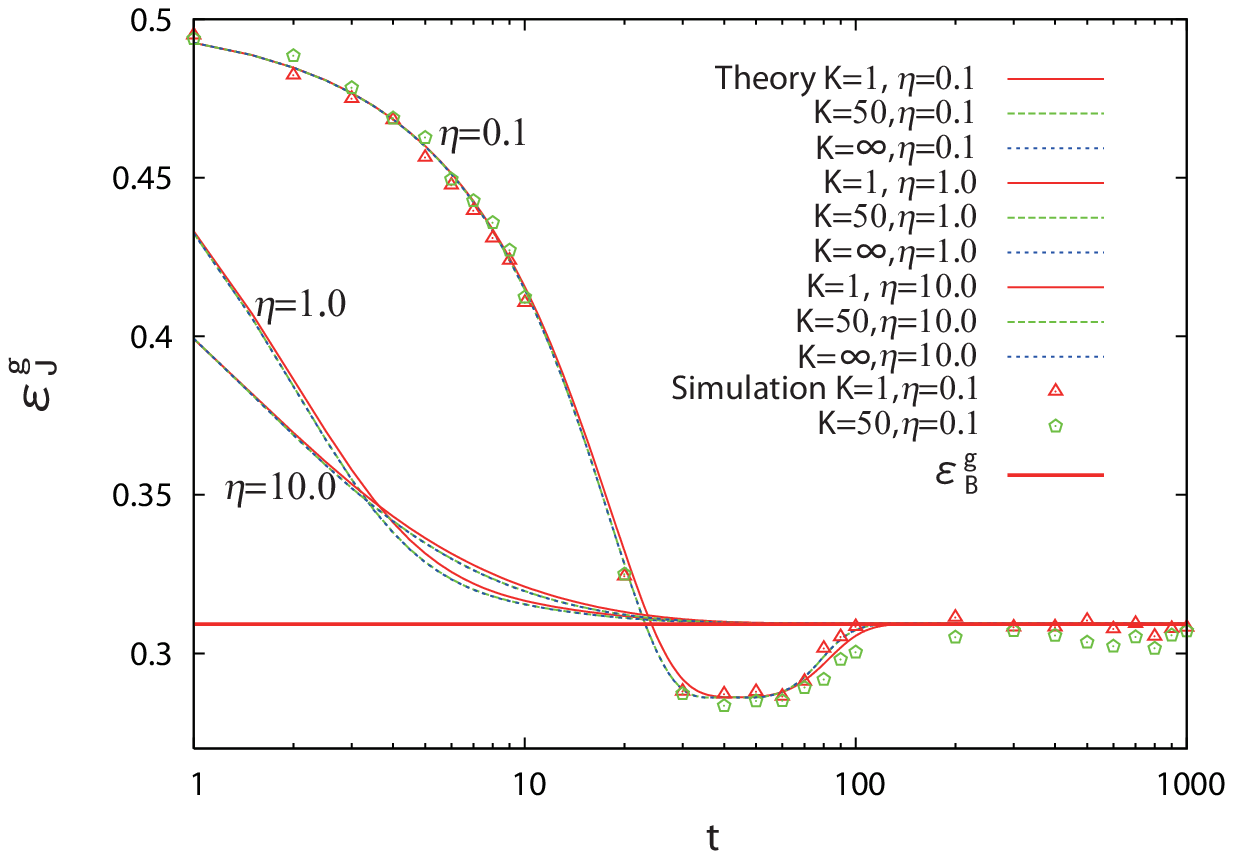}}
     \caption{Time dependence of the generalization error of the student
  $\epsilon_{J}$ between the student \mvec{J} with the perceptron
  learning and the true teacher \mvec{A} with $a=0.5$ in the case of the
  mobile ensemble teachers.  The symbols of the lines and plots are the same as
 in Fig.~\ref{f1}. }
     \label{f8}
 \end{center}
\end{figure}

Fugues~\ref{f7} and \ref{f8} show the corresponding dynamical behavior of
the generalization errors $\epsilon_J^g$ of the perceptron-learning
student with the fixed and mobile ensemble teachers, respectively. As
expected from the behavior  of $R_J$ in Figs.~\ref{f5} and \ref{f6}, the
steady value of $\epsilon_J^g$ for all the case is the same as that of
the ensemble teachers. However, an essential difference is found in
transient behavior of $\epsilon_J^g$. Although the minimum value does not
necessarily achieve  the fundamental minimum value of $\epsilon_J^g$ in
the case of the fixed ensemble teachers, it does for small value of
$\eta$ in the moving ensemble teachers  with a finite time interval as
shown in Fig.~\ref{f8}. 
This means again that moving the ensemble teachers plays an important
role for the learning performance of the student. 


¡¡

\section{Conclusion}
We have analyzed the generalization performance of a student 
supervised by ensemble moving teachers in the framework of on-line learning. 
In this paper we adopted a non-monotonic perceptron as a true teacher and 
a simple perceptron as the ensemble moving teachers and the student. 
We have treated the Hebbian learning and the perceptron learning as a
learning rule for the student
and have calculated the generalization error of the student with some order 
parameters analytically or numerically. In this study, we particularly
focus on the effect of mobile ensemble teachers  on the learning
performance of the student. Therefore, it is assumed that the ensemble teachers  
learn  only from the true teacher by using the perceptron learning and
reach a steady state before the student begins to
learn. This is helpful for separating a transient
learning effect of the ensemble teachers from an intrinsic effect. 

In the Hebbian learning, it has been proven that the number $K$ of the
ensemble teachers is not efficient, but their continuous learning in
their steady state is significantly important for 
the student to come close to the true teacher. In the case that the ensemble teachers 
continue to learn, the value of $R_{J}$ eventually approaches unity, which is independent 
of the learning rate, even if the number $K$ is one. 
Although the student with $R_J=1$ does not always mean a best learning
performance in the Hebbian learning, the minimum value of $\epsilon_J^g$
reaches the fundamental minimum value as a transient state, regardless
of the number $K$. This is sharp contrast to the case of the fixed
ensemble teachers, in which the fundamental minimum value of
$\epsilon_J^g$ never occurs. The time step at which $\epsilon_J^g$ has a
minimum value decreases with increasing  the learning rate $\eta$, but
its precise step has not been predicted theoretically at the present moment.

In the perceptron learning, in contrast to the Hebbian learning, no
significant difference has been found in the steady states. The steady
values of $R_J$ and $\epsilon_J^g$ coincide with those of $R_B$ and
$\epsilon_B^g$ in both of the fixed and mobile ensemble
teachers. However, the effect of the movement of the ensemble teachers
appears in the transient state in the learning process, where,  in
particular for the small value of the learning rate $\eta$, the maximum
value of $R_J$ exceeds the value of $R_B$ and then the minimum value of
$\epsilon_J^g$ reaches the fundamental minimum value even if the number
$K$  is one. In the case of the fixed ensemble teachers, while the
former is found only for the large $K$ and small $\eta$, the latter is
hardly seen for any parameter observed. It would be interesting to see
that the result of the mobile ensemble teachers weakly depends on the
number of the ensemble teachers. Further, the minimum value of
$\epsilon_J^g$ for the $K=1$ mobile ensemble teacher is smaller than
that for $K=\infty$ fixed ensemble teachers. Our study suggests that the
movement of the ensemble teachers, rather than the number $K$,  is
important for the student learning in our model.

One of the drawbacks of the present model is that the minimum of
$\epsilon_J^g$ is given as the transient state in the learning process
and that no algorithm is found to stop the  learning at the transient
state.  We point out that the perceptron learning shows a finite time
interval of the transient state which gives the minimum of $\epsilon_J^g$
as shown in Fig.~\ref{f8}. This might be convenient in comparison to the
Hebbian learning, but the explicit construction of the stopping
algorithm, including a practical way, still remains to be solved in further
work.

\section*{Acknowledgments}
We are grateful to S. Miyoshi for a critical reading of this manuscript
and 
fruitful discussions. This work was supported by the Grant-in-Aid for Scientific
Research on the Priority Area ``Deepening and Expansion of  Statistical
Mechanical Informatics'' (No. 1807004)  by Ministry of Education, Culture, Sports,
Science and Technology.

\appendix
\section{Derivation of the learning dynamics for the ensemble teachers}
\label{sec:A}
In this appendix, we derive a set of the ordinal differential equations
(\ref{eqn:opeL}), (\ref{eqn:opeR}) and (\ref{eqn:opeQ}) of the order
parameters for the ensemble moving teachers in our model.  
From the update rules of the ensemble teachers of
eq.~(\ref{eqn:updateB}), a standard calculus\cite{1}  leads to the following
ordinal differential equations in terms of the average over the
correlated Gaussian variables, 
\begin{align}
\frac{dl_{B_{k}}}{dt'}=&\langle f_{k}v_{B_{k}} \rangle+\frac{\langle f_{k}^2 \rangle}{2l_{B_{k}}}, \label{c1}\\
\frac{dR_{B_{k}}}{dt'}&=-\frac{R_{B_{k}}}{l_{B_{k}}}\frac{dl_{B_{k}}}{dt'}+\frac{\langle f_{k}v \rangle}{l_{B_{k}}}, \label{c2}\\
\frac{dq_{kk'}}{dt'}&=-\frac{q_{kk'}}{l_{B_{k}}}\frac{dl_{B_{k}}}{dt'}-\frac{q_{kk'}}{l_{B_{k'}}}\frac{dl_{B_{k'}}}{dt'} \nonumber \\
& \quad \, +\frac{\langle f_{k'} v_{B_{k}}\rangle}{ l_{B_{k'}}}+\frac{\langle f_k v_{B_{k'}}\rangle}{l_{B_{k}}} +\frac{\langle f_k f_{k'}\rangle}{l_{B_{k'}}l_{B_{k}}} \label{c3},
\end{align}
where the continuous time $t'$ is defined by the thermodynamic limit of
$m'/N$ with $m'$ being the time step of the ensemble teachers in
eq.~(\ref{eqn:updateB}). The bracket $\langle\cdots\rangle$  denotes the
average with respect to the multiple Gaussian distribution 
given in eq.~(\ref{eqn:multiG}).   
Since each component of \textrm{\boldmath $A$} and \textrm{\boldmath
$B$}$_{k}^{0}$ are generated independently from the Gaussian
distribution,  
\textrm{\boldmath $A$} and \textrm{\boldmath $B$}$_{k}^{0}$ with any $k$
are orthogonal to each other in the thermodynamic limit.
Then, the initial conditions of the differential equations for $R_{B_k}$
and $q_{kk'}$ are given by 
\begin{align}
R_{B_{k}}^{0}&=0, \quad q_{kk'}^{0}=0, \label{initial1}
\end{align}
One easily finds that from eqs. (\ref{c1})-(\ref{c3}) and  (\ref{initial1}) that the order parameters $R_{B_{k}}$, $l_{B_{k}}$ and
$q_{kk'}$ 
are invariant under a permutation of the index $k$ of the ensemble
teachers. Because of the symmetry, we omit the subscripts $k$ from the
order parameters.  
We can calculate sample averages in eqs. (\ref{c1})-(\ref{c3}) and
obtain 
\begin{align}
\langle f_{k}v_{B_{k}} \rangle =& \frac{\eta_B}{\sqrt{2\pi}} \left[R_{B}\left\{2\exp{\left(-\frac{a^2}{2}\right)}-1\right\}-1\right],\\
\langle f_{k}^2 \rangle =& 2\eta_B^2\left(\int_{-\infty}^{-a}+\int_{0}^{a}\right)Dv H\left(-\frac{R_{B}v}{\sqrt{1-R_{B}^2}}\right) ,\\
\langle f_{k}v \rangle =& \frac{\eta_B}{\sqrt{2\pi}}\left\{2\exp{\left(-\frac{a^2}{2}\right)}-R_{B}-1\right\}, \\
\langle f_{k'} v_{B_{k}} \rangle =&\langle f_{k} v_{B_{k'}} \rangle =\frac{\eta_B}{\sqrt{2\pi}}\left[R_{B}\left\{2\exp{\left(-\frac{a^2}{2}\right)}-1\right\}-q\right],\\
\langle f_{k} f_{k'} \rangle =&2\eta_{B}^2\left(\int_{-\infty}^{-a}+\int_{0}^{a}\right)Dv\int_{-\frac{R_{B}v}{\sqrt{1-R_{B}^2}}}^{\infty}DxH(z), 
\end{align} 
where 
\begin{equation}
z\equiv-\frac{(q-R_{B}^2)x+R_{B}\sqrt{1-R_{B}^2}v}{\sqrt{(1-q)(1+q-2R_{B}^2)}}.
\end{equation}
Substituting them into eqs.~(\ref{c1}), (\ref{c2}) and (\ref{c3}), the
differential equations (\ref{eqn:opeL}), (\ref{eqn:opeR}) and
(\ref{eqn:opeQ}) are derived. 

\section{Derivation of the learning dynamics for the student}
\label{sec:B}
As in the appendix\ref{sec:A}, a set of the differential equations for
the student dynamics is derived in this appendix. 
From the update rule (\ref{eqn:updateJ}) of the student, the standard
calculus again leads to the following equations: 
\begin{align}
\frac{dl}{dt}&=\frac{1}{K}\sum_{k=1}^{K}\left(\langle g_{k}u \rangle+\frac{\langle g_{k}^2 \rangle}{2l}\right), \label{s1}\\
\frac{dR_{J}}{dt}&=-\frac{R_{J}}{l}\frac{dl}{dt}+\frac{1}{K}\sum_{k=1}^{K}\frac{\langle g_{k}v \rangle}{l},\label{s2}\\
\frac{dR_{B_{k}J}}{dt}&= -\frac{R_{B_{k}J}}{l}\frac{dl}{dt}-\frac{R_{B_{k}J}}{l_{B_{k}}}\frac{dl_{B_{k}}}{dt} \nonumber \\
&+\frac{1}{K}\sum_{k'=1}^{K}\left(\frac{\langle f_k u \rangle}{l_{B_{k}}}+\frac{\langle g_{k'} v_{B_{k}}\rangle}{l}+\frac{\langle f_k g_{k'}\rangle}{l_{B_{k}}l}\right), \label{s3}
\end{align}
where $t$ denotes a continuous time defined by $t=m/N$. 
As an initial condition of eqs.~(\ref{s2}) and (\ref{s3}), we take 
\begin{equation}
R_{J}^{0}=0, \quad R_{B_{k}J}^{0}=0, \label{initial2} 
\end{equation}
since \textrm{\boldmath $A$}, \textrm{\boldmath $B$}$_{k}^{0}$ and
\textrm{\boldmath $J$}$^{0}$ are orthogonal to each other in the
thermodynamic limit. It is shown from eqs. (\ref{initial2}) and
(\ref{s3}) that the order parameter $R_{B_{k}J}$ does not depend on the
index $k$. Then, one can omit the subscript $k$ from the order parameter
without loss of the generality.  By substituting the two update
functions $g$ of the Hebbian and the perceptron learning respectively,
one calculates the Gaussian averages in
eqs. (\ref{s1})-(\ref{s3})  
in the case of the Hebbian learning as 
\begin{align}
\langle g_{k}u \rangle&=\eta\sqrt{\frac{2}{\pi}}R_{BJ}, \label{eqn:Hebb-b}\\
\langle g_{k}^2 \rangle&=\eta^2 ,\\
\langle g_{k}v \rangle &= \eta\sqrt{\frac{2}{\pi}}R_{B}, \\
\langle f_{k}u \rangle&=\frac{\eta_B}{\sqrt{2\pi}}\left[R_{J}\left\{2\exp{\left(-\frac{a^2}{2}\right)}-1\right\}-R_{BJ}\right],\\
\langle g_{k'}v_{B_{k}}\rangle&= \eta\sqrt{\frac{2}{\pi}}q\delta_{k,k'}, \\
\langle f_{k}g_{k'} \rangle&=-2\eta\eta_B\left[\left(\int_{-\infty}^{-a}+\int_{0}^{a}\right)Dv \int_{-\frac{R_{B}v}{\sqrt{1-R_{B}^2}}}^{\infty} Dx \left\{2H(z)-1\right\}\right], \\
\langle f_{k}g_{k} \rangle&=
 -2\eta\eta_B\left(\int_{-\infty}^{-a}+\int_{0}^{a}\right)Dv
 H\left(-\frac{R_{B}v}{\sqrt{1-R_{B}^2}}\right), \label{eqn:Hebb-f}
\end{align}
and in the case of the perceptron learning as 
\begin{align}
\langle g_{k}u \rangle&=\frac{\eta}{\sqrt{2\pi}}(R_{BJ}-1), \label{eqn:percep-b}\\
\langle g_{k}^2 \rangle&=\frac{\eta^2}{\pi}\tan^{-1}\left(\frac{\sqrt{1-R_{BJ}^2}}{R_{BJ}}\right),\\
\langle g_{k}v \rangle &= \frac{\eta}{\sqrt{2\pi}}(R_{B}-R_{J}), \\
\langle f_{k}u \rangle&=\frac{\eta_B}{\sqrt{2\pi}}\left[R_{J}\left\{2\exp{\left(-\frac{a^2}{2}\right)}-1\right\}-R_{BJ}\right],\\
\langle g_{k'}v_{B_{k}}\rangle&= \frac{\eta}{\sqrt{2\pi}}(q\delta_{k,k'}-R_{BJ}), \\
\langle f_{k}g_{k'} \rangle&=2\eta\eta_B \left(\int_{-\infty}^{-a}+\int_{0}^{a}\right)Dv \int_{-\frac{R_{B}v}{\sqrt{1-R_{B}^2}}}^{\infty} Dx \left\{-\int_{z}^{\infty}DyH\left(-z_1\right)+\int_{-\infty}^{z}DyH\left(z_1\right)\right\}, \nonumber \\
&&\\
\langle f_{k}g_{k}
 \rangle&=2\eta\eta_B\left(\int_{-\infty}^{-a}+\int_{0}^{a}\right)Dv
 \int_{-\frac{R_{B}v}{\sqrt{1-R_{B}^2}}}^{\infty} Dx
 \left\{2H(z_2)-1\right\}. \label{eqn:percep-f}
\end{align}
Here, $z_{1}$ and $z_{2}$ are defined as 
\begin{equation}
z_{1}\equiv-\frac{(R_{BJ}-R_{B}R_{J})\left(\sqrt{1-q}y+\sqrt{1+q-2R_{B}^2}x\right)+R_{J}\sqrt{(1-R_{B}^2)(1+q-2R_{B}^2)}v}{\sqrt{(1-R_{B}^2)\left\{(1+q)(1-R_{J}^2)-2(R_{B}^2-2R_{B}R_{J}R_{BJ}+R_{BJ}^2)\right\}}} 
\end{equation}
and 
\begin{equation}
z_{2}\equiv-\frac{(R_{BJ}-R_{B}R_{J})x+R_{J}\sqrt{1-R_{B}^2}v}{\sqrt{1-R_{J}^2-R_{B}^2-R_{BJ}^2+2R_{B}R_{J}R_{BJ}}},  
\end{equation}
and $\delta_{k,k'}$ is the Kronecker delta defined by 
\begin{equation}
\delta_{k,k'}=\left\{
\begin{array}{ll}
 +1, &\quad k=k',  \\
 \,\,\,\,0, &\quad k\neq k'.
\end{array}
\right.
\end{equation}
Inserting (\ref{eqn:Hebb-b})-(\ref{eqn:Hebb-f}) and
(\ref{eqn:percep-b})-(\ref{eqn:percep-f}) into (\ref{s1})-(\ref{s3})
gives the dynamical equations (\ref{eqn:opeSH1})-(\ref{eqn:opeSH3}) for
the Hebbian rule and those (\ref{eqn:opeSP1})-(\ref{eqn:opeSP3}) for the
perceptron one, respectively.

\end{document}